\newcommand{\sfrac}[2]{{\textstyle{#1\over#2}}}
\documentstyle{article}
\begin{document}
\title{Suppression of gravitational radiation}
\author{W.B.Bonnor and M.S.Piper
                   \\ School of Mathematical Sciences, \\
                   Queen Mary and Westfield College,\\
                   Mile End Road,\\ 
                   London.\\
                   E1 4NS}
\maketitle
\setlength{\parindent}{0.5in}
\begin{abstract}
We consider a burst of quadrupole gravitational radiation
in the presence of a large static mass $M$ situated at its source.
Some of the radiation is back-scattered off the static field of the large mass,
forming a wave tail. After the burst, the tail is a pure incoming wave,
carrying energy back towards the source.  We calculate this energy, and, in a
numerical example,
compare it with the outgoing wave energy.  If $M$ is sufficiently large
the incoming energy can equal the outgoing energy, indicating that the primary
outgoing wave is completely suppressed.
\end{abstract}

\section{Introduction}
It has been known for many years that gravitational waves have tails [1][2][3].
(For a good literature survey see [4]).
Tails are now recognised as factors in the planned observational
detection of the waves [5][14][15].  The simplest sort of tail results from
the back-scattering of an outgoing wave by the static field of the source,
or of other bodies.  Let us suppose for simplicity that there is an outgoing
quadrupole wave that lasts for a finite time; when it  ceases the tail 
persists, and as shown by Hunter and Rotenberg [3], it is then a pure incoming 
quadrupole wave.
As such it carries energy, and this is what we calculate in this paper.  The
fate of this energy is not certain.  It might be reabsorbed by the source, it
might pass through the centre and become outgoing, or it might even form a
singularity as suggested by the numerical work of Abraham and Evans[6][7].
In this paper we do not consider these possibilities, but merely study the
relative energies of the primary outgoing wave and the tail.

The incoming energy carried by the tail must, of course, decrease the
outgoing wave energy, as given, for example, by the quadrupole formula.
The question therefore arises whether the calculated incoming energy could be 
greater
than the outgoing, which would presumably mean that the static fields of
the source and its neighbours suppress the waves entirely.  From the
theoretical point of view there seems no objection to this, since
a similar suppression can occur of a light wave by a black hole.  Indeed
the suppression was investigated some years ago from a different point of view
by Kundu [8], and his work caused some debate [9][10].  The conclusion seemed
to be that a suppressive  effect existed but would be counteracted by other 
phenomena,
and would have no astrophysical importance [11].

The plan of the paper is as follows.  We describe the metric and approximation
method in Sections 2 and 3, and follow these with the linear approximation
to the outgoing quadrupole wave in Section 4.  We give the formulae for
the wave tail in Section 5, and for the energy it carries after the
vibration ceases in Section 6.  Section 7 contains numerical calculations of
the energies carried outward by the quadrupole wave and inward by its tail
for a specified pulse.  Conclusions are expressed in Section 8.

\section{Metric}
When dealing with outgoing waves
we shall use the axially symmetric metric of Bondi [12]:
\begin{equation}
ds^{2}=-r^{2}(Bd\theta^{2}+B^{-1}\sin^{2}\theta d\phi^{2})+D du^{2} +2F drdu
+2rG d\theta du,
\end{equation}
where $B,D,F,G$ are functions of $r,\theta$ and $u$.  $u$ is a retarded time
coordinate and may be thought of as ordinary time $t$ at the source, which is
represented by a singularity at the origin.  The Schwarzschild solution in the 
coordinates of (1) is
\begin{equation}
ds^{2}=-r^{2}(d\theta^{2}+\sin^{2}\theta d\phi^{2})+(1-2Mr^{-1})du^{2} + 2drdu.
\end{equation}

We shall also be studying incoming waves and for these we use the Bondi metric
in the form
\begin{equation}
ds^{2}=-r^{2}(Bd\theta^{2} +B^{-1}\sin^{2}\theta d\phi^{2})+Ddv^{2}-2Fdrdv
+2rG d\theta dv,
\end{equation}
with $B,D,F,G$ now functions of $r,\theta$ and $v$, and $v$ an advanced time 
coordinate. In this metric the Schwarzschild solution is
\begin{equation}
ds^{2}=-r^{2}(d\theta^{2}+\sin^{2}\theta d\phi^{2})+(1-2Mr^{-1})dv^{2}-2drdv.
\end{equation}
In these metrics the ranges are
\[r>0,\;\;0\leq\theta\leq\pi,\;\;0\leq\phi\leq2\pi,\;\;-\infty<u,v<\infty.\]

\section{The approximation method}
We use an adaptation of the double series method.  We suppose that the
solution we are looking for can be expanded in a doubly infinite power
series in $M$, a central spherical mass, and a parameter $Q$ characterising
the primary\footnote{By primary quadrupole moment is meant the one inserted
at the linear approximation, generating waves against a Minkowski background, 
and entering the usual quadrupole formula.  Quadrupole waves appear in higher 
approximations
and one might regard them as arising from higher quadrupole moments, though
we do not do this here. Hereafter we drop the prefix `primary', and refer 
simply to the quadrupole moment.}
quadrupole moment of the vibrating source.  The actual quadrupole moment
at time $t$ is $Qh(t)$, and $h$ is zero except in $t_{1}<t<t_{2}$.  Therefore
we envisage an expansion of the $g_{ik}$ of the form
\begin{equation}
g_{ik}=\sum_{p=0}^{\infty}\sum_{s=0}^{\infty}M^{p}Q^{s}\stackrel{(ps)}{g_{ik}},
\end{equation}
the $\stackrel{(ps)}{g_{ik}}$ being independent of $M$ and $Q$.

If we substitute (5) into the vacuum field equations
\begin{equation}
R_{ik}=0
\end{equation}
and consider the coefficient of $M^{p}Q^{s}$ we find seven equations of the
form
\begin{equation}
\Phi_{lm}(\stackrel{(ps)}{g_{ik}})=\Psi(\stackrel{(qr)}{g_{ik}}),
\end{equation}
where the left-hand side is linear in the $\stackrel{(ps)}{g_{ik}}$ (and
their derivatives), and the right-hand side consists of terms non-linear
in the $\stackrel{(qr)}{g_{ik}}$ and their derivatives, with 
$q \leq p,r \leq s$ and
$q+r<p+s$.  This is called the $(ps)$ approximation.
We can introduce this notation into the metric coefficients of (1), for
example:
\[-g_{22}=r^{2}B=r^{2}[1+
\sum_{p=0}^{\infty}\sum_{s=0}^{\infty}M^{p}Q^{s}\stackrel{(ps)}{B}]\]
where $\stackrel{(00)}{B}$ is zero.
In this paper we shall
go in detail only as far as the $(11)$ approximation; this will express the
interaction of the central mass and the quadrupole oscillations, which
describes the simplest wave tail.  To calculate the energies
transmitted by the quadrupole wave and its tail we shall need formulae from 
the $(02)$
and $(22)$ approximations respectively which are known from previous work,
namely eqns (12) and (29).

\section{The linear approximation}
We write down the solution, in the coordinates of (1), of the linearised form
of (6) (i.e. (7) with $\Psi=0$) corresponding to a outgoing quadrupole
wave [1].
\begin{eqnarray}
\stackrel{(01)}{B}&=&\frac{1}{2} \sin^{2}\theta(r^{-1}\ddot{h}+r^{-3}h),\\
\stackrel{(01)}{D}&=&-P_{2}(2r^{-1}\ddot{h}+2r^{-2}\dot{h}+r^{-3}h),\\
\stackrel{(01)}{F}&=&0,\\
\stackrel{(01)}{G}&=&\frac{1}{2}\sin\theta\cos\theta(-2r^{-1}\ddot{h}
+4r^{-2}\dot{h}+3r^{-3}h),
\end{eqnarray}
where $P_{2}(\cos\theta)$ is the Legendre polynomial and an overdot means 
differentiation with respect to $u$.  $h(u)$ is zero except for
$u_{1}<u<u_{2}$, where $u_{i}=t_{i}$, and represents the quadrupole 
oscillation.  We write
$Qh$ for the quadrupole moment of the source.   We shall assume $h$ is of
class $C^{3}$ for $u>u_{1}$ so that we can perform the necessary calculations
on the tail in Section 7.

The energy carried outward by this oscillation, calculated by the usual
quadrupole formula or as a loss of mass as in [1], is
\begin{equation}
E_{out}=\frac{Q^{2}}{30}\int_{u_{1}}^{u_{2}}\stackrel{...}{h}^{2}du.
\end{equation}

\section{The wave tail}
The tail comes from the interaction of the quadrupole wave with the 
Schwarzschild
metric.  These are respectively the $(01)$ and the $(10)$ approximations, given
by (8)-(11) and (2).  Inserting them into the right-hand side of (7) we have
seven differential equations for the $\stackrel{(11)}{g_{ik}}$, of which the
solution is [3][13]. \footnote{In the previous work cited the notation
was different and the tail approximation was denoted $(22)$ instead of $(11)$.}
\begin{eqnarray}
\stackrel{(11)}{B}&=&\frac{1}{8}\sin^{2}\theta(2r^{-3}\dot{h}+10r^{-4}h+r^{-1}
\frac{\partial^{2} H}{\partial r^{2}}-4r^{-2}\frac{\partial H}{\partial r}
+4r^{-3}H),\\
\stackrel{(11)}{D}&=&-\frac{1}{2}P_{2}(r^{-3}\dot{h}+2r^{-4}h+2r^{-3}H),\\
\stackrel{(11)}{F}&=&0,\\
\stackrel{(11)}{G}&=&\frac{1}{8}\sin2\theta(3r^{-3}\dot{h}+8r^{-4}h
-2r^{-2}\frac{\partial H}{\partial r}
+6r^{-3}H),
\end{eqnarray}
where H(r,u) is a tail function defined by
\begin{equation}
H=-2\frac{\partial}{\partial u}\int_{-\infty}^{u}\frac{h(\xi)d\xi}{(u+2r-\xi)}.
\end{equation}

{}From the definition of $H$ we find
\[\frac{\partial H}{\partial r}-2\frac{\partial H}{\partial u}
=2r^{-1}\dot{h};\]
now $\dot{h}=0$ after the end of the quadrupole oscillation  so
\begin{equation}
H(r,u)=H(u+2r),\;\; u>u_{2}.
\end{equation}
In the flat background spacetime $u+2r$ is the advanced time coordinate
(hereafter denoted by $v$) so {\em for $u>u_{2}$  the tail function $H$
is an incoming wave}.  From (17)
\[H=-\frac{h(u)}{r} +2\int_{-\infty}^{u}\frac{h(\xi)d\xi}{(u+2r-\xi)^{2}}\]
so as $h$ as well as $\dot{h}$ is zero for $u>u_{2}$, we have for
the tail function after the vibration has finished
\begin{equation}
H_{0}=2\int_{-\infty}^{u}\frac{h(\xi)d\xi}{(u+2r-\xi)^{2}},\;u>u_{2}.
\end{equation}

Since $H_{0}$ is a function of $v=u+2r$ we can write
(13)-(16) for $u>u_{2}$ as
\begin{eqnarray}
\stackrel{(11)}{B}&=&\frac{1}{2}\sin^{2}\theta(r^{-1}H_{0}^{\prime\prime}
-2r^{-2}H_{0}^{\prime}+r^{-3}H_{0}),\\
\stackrel{(11)}{D}&=&-P_{2}r^{-3}H_{0},\\
\stackrel{(11)}{F}&=&0,\\
\stackrel{(11)}{G}&=&\frac{1}{4} \sin2\theta (-2r^{-2}H_{0}^{\prime}+
3r^{-3}H_{0}),
\end{eqnarray}
where $\prime$ means $d/dv$.   {\em This is the linear approximation for an
incoming quadrupole wave in outgoing Bondi coordinates}.  It was given in [3],
eqn (7.9) with $s=2$
and $\stackrel{2}{f}=-H_{0}$.  As $v=u+2r$ and $u>u_{2},\;r>0$, (20)-(23)
apply for
\begin{equation}
v>u_{2},
\end{equation}
$u_{2}$ being, of course, a constant.

In order to calculate the energy carried by this incoming wave we write it
in {\em incoming} Bondi coordinates, eqn (3).  This too was done in [3] eqn 
(A25),
and the result is
\begin{eqnarray}
\stackrel{(11)}{B}&=&\frac{1}{2}\sin^{2}\theta(r^{-1}H_{0}^{\prime\prime}+
r^{-3}H_{0}),\\
\stackrel{(11)}{D}&=&-P_{2}(2r^{-1}H_{0}^{\prime\prime}-2r^{-2}H_{0}^{\prime}+
r^{-3}H_{0}),\\
\stackrel{(11)}{F}&=&0,\\
\stackrel{(11)}{G}&=&\frac{1}{2}\sin\theta 
\cos\theta(2r^{-1}H_{0}^{\prime\prime}+4r^{-2}H_{0}^{\prime}-3r^{-3}H_{0}).
\end{eqnarray}
\section{Energy carried by the wave tail}
In the case of outgoing waves the energy carried is given by the loss of 
mass of the
source.  The simplest approximation in which it appears is that of the
quadrupole-quadrupole interaction, so that in our notation of Section 3 it is
of order $Q^{2}$.  This approximation was completely solved in [3], and the
loss of mass, given by eqn (12), agreed with the ordinary quadrupole formula.

The incoming waves described by (25)-(28) are of order $MQ$, but they have
quadrupole form, and one can adapt the calculations of [3] to solve the 
$M^{2}Q^{2}$
approximation in the incoming case.  In fact the solution has exactly the same
form as that of [3] \footnote{There described as the
quadrupole-quadrupole approximation}, except for changes of sign. 
(See Appendix 1).  The result
is that {\em the energy flowing inward after the vibration ceases is}
\begin{equation}
E_{in}=\frac{M^{2}Q^{2}}{30}\int_{u_{2}}^{\infty}H_{0}'''^{2}dv,
\end{equation}
$H_{0}$ being defined by (19).  The lower limit of the integral arises from 
(24).
\section{An example}
To get a feeling for the magnitudes involved we construct an artificial
example in which $h(u)$ is a single pulse given by
\begin{equation}
h(u)=(1-u)^{5},\;-1<u\leq1;\;\;h(u)=0,\;\mid u\mid\geq1.
\end{equation}
This has been chosen to make the tail function $H$ reasonably simple.  Although
$h$ is discontinuous at $u=-1$ it is very smooth at $u=1$ which is sufficient 
to
give an $H_{0}$ which can be used in (29) if we put $u_{2}=1$.  In this way 
we can calculate the
energy carried inward by the tail after the end $u=u_{2}$ of the vibration.

First we calculate the energy $E_{out}$ carried {\em outward} by the pulse $h$ 
according
to (12):
\begin{equation}
E_{out}=\frac{Q^{2}}{30} \int_{-1}^{1}\stackrel{...}{h}^{2}du=768Q^{2}.
\end{equation}
where we have inserted the form (30) for $h$.

{}From (19) we have for $H_{0}(v),\; v\geq1$:
\begin{equation}
H_{0}(v)=2\int_{-1}^{1}\frac{(1-\xi)^{5}}{(v-\xi)^{2}}d\xi,\: v>1.
\end{equation}
whence, after a fairly long but straightforward calculation,
\begin{equation}
H^{\prime\prime\prime}=16(1+v)^{-4}(64 +30v-70v^{2}-90v^{3}-30v^{4})+240(v-1)
[\ln(v+1)-\ln(v-1)].
\end{equation}

To find $E_{in}$ we have to calculate the integral (29). This can be evaluated
exactly using REDUCE.  The result is
\begin{equation}
E_{in}=28.2M^{2}Q^{2}
\end{equation}
to three significant figures.

These expressions for the inward and outward energies are in relativistic 
units.
We need to find their ratio when they are expressed in units of customary
dimensions.  Denoting them in such units by $W_{in}$ and $W_{out}$ we show
in Appendix 2 that the ratio is
\begin{equation}
\frac{W_{in}}{W_{out}}=\frac{G^{2}m^{2}}{c^{6}}
\frac{\int_{1}^{\infty}H_{0}'''^{2}dv}
{\int_{-1}^{1}h'''^{2}du}.
\end{equation}
Using the values of the integrals calculated above we find
\begin{equation}
\frac{W_{in}}{W_{out}}\sim 2\times10^{-79}m^{2},
\end{equation}
where $m$ is in grams.  To render $W_{in}$ comparable with $W_{out}$ the
central mass would have to be that of a large black hole.

\section{Conclusion}
We have shown that in the presence of a sufficiently large mass the incoming
energy of the tail can be comparable or even equal to that of the outgoing
wave pulse.  This raises the possibility that a pulse of radiation from, say,
a coalescing binary might be significantly reduced, or even suppressed, if
the static gravitational fields in its neighbourhood are great enough.

It must be emphasised that our calculations are not intended as a serious
astrophysical model.  We have treated the neighbours of the source
as a single static spherical mass at the position of the source, and our 
numerical calculations are
for one very special shape of pulse.  Moreover, we have considered the tail 
only
after the primary pulse has ceased.  Nevertheless, we believe that the
incoming tail energy should be taken carefully into account in
gravitational wave experiments.

\section{Appendix 1}
In this Appendix we derive formulae relating to the incoming wave-tail energy
referred to in Section 6.
For this we make use of the outgoing quadrupole-quadrupole solution given by
equations (3.7)-(3.11) in [3]. To obtain the complete $M^{2}Q^{2}$ solution
in the incoming case we simply make the coordinate transformation
\begin{equation}
u=-v
\label{uv}
\end{equation}
and the identification
\begin{equation}
h=H_{0}
\end{equation}
with $r$, $\theta$ and $\phi$ remaining unchanged and where $h$ is written as 
$\stackrel{2}{h}$ in [3]. The only part of the transformation
 that is not entirely straightforward is the way in which
$\stackrel{(24)}{Y}$ of [3] transforms to its equivalent $Y$ here. The former
is defined by 
\begin{equation}
\frac{d\stackrel{(24)}{Y}}{du}=\stackrel{...}{h}^{2}
\end{equation}
where and overdot denotes differentiation with respect to $u$,
and so $Y$ must satisfy
\begin{equation}
-\frac{dY}{dv}=H_{0}'''^{2}.
\end{equation}
where $^{'}$ denotes differentiation with respect to $v$.
Integrating with respect to $v$ gives
\begin{equation}
Y=-\int_{u_{2}}^{\infty}H_{0}'''^{2}dv.
\end{equation}
The transformation (\ref{uv}), used on (3.7)-(3.11) of [3] then gives:
\begin{eqnarray}
\stackrel{(22)}{D}&=&[(\sfrac{3}{5}-3s^{2}+\sfrac{21}{8}s^{4})
H_{0}'H_{0}^{\mbox{{\tiny IV}}}+(\sfrac{12}{5}-9s^{2}+\sfrac{27}{4}s^{4})
H_{0}''H_{0}'''+\sfrac{1}{15}Y]r^{-1}\nonumber \\ 
& &+[(-2+10s^{2}-\sfrac{35}{4}s^{4})H_{0}'H_{0}'''+(-2+9s^{2}
-\sfrac{15}{2}s^{4})H_{0}''^{2}]r^{-2}\nonumber \\
& &+(\sfrac{9}{2}-\sfrac{47}{2}s^{2}+\sfrac{331}{16}s^{4})H_{0}'
H_{0}''r^{-3}+[(\sfrac{3}{2}s^{2}-\sfrac{27}{16}s^{4})H_{0}H_{0}''\nonumber \\
& &+(-\sfrac{15}{4}+\sfrac{71}{4}s^{2}-\sfrac{493}{32}s^{4})H_{0}'^{2}]
r^{-4}+(\sfrac{3}{2}-6s^{2}+\sfrac{81}{16}s^{4})H_{0}'H_{0}r^{-5}\nonumber \\
& &+(\sfrac{1}{2}-3s^{2}+\sfrac{21}{8}s^{4})H_{0}^{2}r^{-6}\\
\stackrel{(22)}{F}&=&s^{4}[-\sfrac{1}{32}H_{0}''^{2}r^{-2}-\sfrac{3}{32}H_{0}
H_{0}''r^{-4}-\sfrac{3}{32}H_{0}^{2}r^{-6}]\\
\stackrel{(22)}{G}&=&[(-\sfrac{3}{10}cs+\sfrac{21}{40}cs^{3})H_{0}'
H_{0}^{\mbox{{\tiny IV}}}+(-\sfrac{6}{5}cs+\sfrac{27}{20}cs^{3})H_{0}''
H_{0}'''-(\sfrac{1}{30}cs+\sfrac{1}{40}cs^{3})Y]r^{-1}\nonumber \\
& &+[(-2cs+\sfrac{7}{2}cs^{3})H_{0}'H_{0}'''+(-2cs+
\sfrac{23}{8}cs^{3})H_{0}''^{2}]r^{-2}\nonumber \\
& &+(\sfrac{27}{4}cs-\sfrac{185}{16}cs^{3})H_{0}'H_{0}''r^{-3}+
[\sfrac{1}{8}cs^{3}H_{0}H_{0}''+(-\sfrac{15}{2}cs+\sfrac{105}{8}cs^{3})
H_{0}'^{2}]r^{-4}\nonumber \\
& &+(\sfrac{15}{4}cs-\sfrac{107}{16}cs^{3})H_{0}H_{0}'r^{-5}+
(\sfrac{3}{2}cs-\sfrac{21}{8}cs^{3})H_{0}^{2}r^{-6}\\
\stackrel{(22)}{B}&=&[(-\sfrac{3}{20}s^{2}+\sfrac{7}{40}s^{4})
H_{0}'H_{0}^{\mbox{{\tiny IV}}}+(-\sfrac{3}{5}s^{2}+\sfrac{9}{20}s^{4})
H_{0}''H_{0}'''+(-\sfrac{1}{60}s^{2}-\sfrac{1}{120}s^{4})Y]r^{-1}
\nonumber \\
& &+\sfrac{1}{8}s^{4}H_{0}''^{2}r^{-2}+(-\sfrac{9}{4}s^{2}+
\sfrac{21}{8}s^{4})H_{0}'H_{0}''r^{-3}+[\sfrac{1}{4}s^{4}H_{0}H_{0}''+
(\sfrac{75}{16}s^{2}-\sfrac{175}{32}s^{4})H_{0}'^{2}]r^{-4}\nonumber \\
& &+(-\sfrac{27}{8}s^{2}+\sfrac{69}{16}s^{4})H_{0}'H_{0}r^{-5}+
(-\sfrac{7}{4}s^{2}+\sfrac{15}{8}s^{4})H_{0}^{2}r^{-6}\\
\stackrel{(22)}{C}&=&[(\sfrac{3}{20}s^{2}-\sfrac{7}{40}s^4)
H_{0}'H_{0}^{\mbox{\tiny IV}}+(\sfrac{3}{5}s^2-\sfrac{9}{20}s^4)
H_{0}''H_{0}'''+(\sfrac{1}{60}s^2+\sfrac{1}{120}s^4)Y]
r^{-1}\nonumber \\
& &+\sfrac{1}{8}s^4H_{0}''^2r^{-2}+(\sfrac{9}{4}s^2-\sfrac{21}{8}s^4)
H_{0}'H_{0}''r^{-3}+[\sfrac{1}{4}s^4H_{0}H_{0}''+(-\sfrac{75}{16}s^2+
\sfrac{175}{32}s^4)H_{0}'^2]r^{-4}\nonumber \\
& &+(\sfrac{27}{8}s^2-\sfrac{69}{16}s^4)H_{0}'H_{0}r^{-5}+(\sfrac{7}{4}s^2-
\sfrac{13}{8}s^4)H_{0}^2r^{-6}
\end{eqnarray}
where $s\equiv\sin\theta$ and $c\equiv\cos\theta$.
By a coordinate transformation equivalent to (9.13) in [1], it can be shown 
that the above (22) solution describes, after the vibration, energy flowing 
inward equal to
\begin{equation}
E_{in}=\frac{M^{2}Q^{2}}{30}\int_{u_{2}}^{\infty}H_{0}'''^{2}dv.
\end{equation}
\section {Appendix 2}
We consider here the ratio of energies transmitted inward by the tail and
outward by the primary quadrupole oscillation discussed in Section 7.

We now write the energies in units of customary dimensions (ucd) and denote 
them by
$W_{out}$ and $W_{in}$.  From (12)
\begin{equation}
W_{out}=\frac{GQ^{2}}{30c^{5}}\int_{u_{1}}^{u_{2}}\stackrel{...}{h}^{2}du.
\end{equation}
To express $W_{in}$ we first note that the dimensions of $H$ are
(time)$^{-1}\times$ (dimensions of $h$), so in ucd $H_{0}$ becomes
\[\frac{2}{c}\int_{-\infty}^{u}\frac{h(\xi)d\xi}{(v-\xi)^{2}}.\]
We must also remember that $M$ must be replaced by $Gm/c^{2}$, where $m$ is
the central mass in ucd.  Using this we have from (29)
\begin{equation}
W_{in}=\frac{G}{c^{5}}(\frac{Gm}{c^{2}})^{2}\frac{Q^{2}}{30}
\frac{1}{c^{2}}\int_{u_{2}}^{\infty}H_{0}'''^{2}dv.
\end{equation}
Eqn (35) follows from (37) and (38) if we \vspace {0.2in} put $u_{1}=-1, 
u_{2}=1.$\\
{\sc\large REFERENCES}\\
{[1]}  Bonnor W B and Rotenberg M A 1966 {\em Proc. R. Soc.} A{\bf289} 247\\
{[2]}  Couch W E, Torrence R J, Janis A I and Newman E T 1968 {\em J. Math. 
Phys.} {\bf9} 484\\
{[3]}  Hunter A J and Rotenberg M A 1969 {\em J. Phys. A (Gen. Phys.)} {\bf2} 
34\\
{[4]}  Blanchet L and Sch\"{a}fer G 1993 {\em Class. Quantum Grav.} {\bf10} 
2699\\
{[5]}  Blanchet L and Sathyaprakash B S 1994 {\em Class. Quantum Grav.} 
{\bf11} 2807\\
{[6]}  Abrahams A M and Evans C R 1992 {\em Phys. Rev.} D{\bf46} R4117\\
{[7]}  Abrahams A M and Evans C R 1993 {\em Phys. Rev. Lett.} {\bf70} 2980\\
{[8]}  Kundu P K 1990 {\em Proc. R. Soc.} A{\bf431} 337\\
{[9]}  Price R H and Pullin J 1992 {\em Phys. Rev.} D{\bf46} 2497\\
{[10]} Kozameh C, Newman E T and Rovelli C 1991 {\em Phys. Rev.} D{\bf44} 
551\\
{[11]} Price R H, Pullin J and Kundu P K 1993 {\em Phys. Rev. Lett.} {\bf70} 
1572\\
{[12]} Bondi H, van der Burg M G J and Metzner A W K 1962 {\em Proc. R. Soc.} 
A{\bf269} 21\\
{[13]} Bonnor W B 1974 {\em Ondes et Radiations Gravitationelles} (Paris:CNRS)
p73\\
{[14]} Blanchet L and Damour T 1992 {\em Phys. Rev.} D{\bf46} 4304\\
{[15]} Blanchet L 1997 {\em Phys. Rev.} D{\bf55} 714

\end{document}